# Monte Carlo study of 8-state Potts model on 2D random lattices [*]

Wolfhard Janke[a][†] and Ramon Villanova[b]

[a]Institut für Physik, Johannes Gutenberg-Universität Mainz,
Staudinger Weg 7, 55099 Mainz, Germany

[b]Matematiques Aplicades, Universitat Pompeu Fabra,
La Rambla 32, 08002 Barcelona, Spain

We study the effect of quenched coordination-number disorder of random lattices on the nature of the phase transition in the two-dimensional eight-state Potts model, which is of first order on regular lattices. We consider Poissonian random lattices of toroidal topology constructed according to the Voronoi/Delaunay prescription. Monte Carlo simulations yield strong evidence that the phase transition remains first order.

## 1. INTRODUCTION

Pure systems exhibiting a continuous phase transition are very susceptible to the addition of random disorder. The critical behaviour can be driven to new universality classes or the phase transition can be eliminated altogether [1]. Also for first-order phase transitions phenomenological renormalization-group arguments suggest strong effects caused by random disorder [2]. In particular the order of the transition can change from first to second.

The well-known paradigm to investigate such effects is the two-dimensional $q$-state Potts model which undergoes on regular lattices for $q \geq 5$ a temperature driven first-order phase transition [3]. Monte Carlo (MC) simulations for $q = 8$ with a certain type of quenched bond randomness provided clear evidence for a continuous phase transition of the Ising type [4]. Also in two-dimensional quantum gravity studies of Potts "matter" coupled to dynamically triangulated random surfaces (DTRS) a similar softening effect was observed [5]. From a statistical mechanics viewpoint, in this case the Potts model is subject to annealed disorder in the local coordination numbers of the dynamical triangulation.

[*]Work supported in part by NATO collaborative research grant CRG940135.
[†]WJ thanks the Deutsche Forschungsgemeinschaft for a Heisenberg fellowship, and also acknowledges support in part by EEC grant ERBCHRXCT930343.

Here we report on a study [6] of the same model on static Poissonian random lattices constructed according to the Voronoi/Delaunay prescription [7]. The locally varying coordination numbers cause the disorder similar to Ref. [5], but in our case the disorder is assumed to be frozen in, i.e. "quenched", as in Ref. [4].

## 2. MODEL AND SIMULATION

The 8-state Potts model is defined by the partition function

$$Z = \sum_{\{\sigma_i\}} e^{-\beta E}; E = -\sum_{\langle ij \rangle} \delta_{\sigma_i \sigma_j}; \sigma_i = 1, \ldots, q, \quad (1)$$

with $q = 8$. The $\sigma_i$ are integer valued spins at the lattice sites $i$, $\delta_{\sigma_i \sigma_j}$ denotes the usual Kronecker delta symbol, and the nearest-neighbor bonds $\langle ij \rangle$ are determined by the Voronoi/Delaunay construction of the random lattices. We always used periodic boundary conditions, i.e., toroidal topology as depicted in Fig. 1.

Using a standard algorithm [7,8] we generated 20 independent replica of random lattices with $V = 250, 500, 750, 1000, 2000$, and $3000$ sites and performed long single-cluster simulations near the transition point at $\hat{\beta} = 0.826, 0.830, 0.830, 0.830, 0.832$, and $0.833$, respectively. After equilibration we recorded 1 000 000 measurements (taken after 1, 1, 1, 1, 2, 4 clusters had been flipped) of the energy $E$ and the magnetization $M = (q \max\{n_i\} - V)/(q-1)$ in a time-series file, where $n_i \leq V$ denotes the number of spins of "orientation" $i = 1, \ldots, q$ in one lattice configuration. The corresponding quantities per site will be denoted by $e = E/V$ and $m = M/V$.

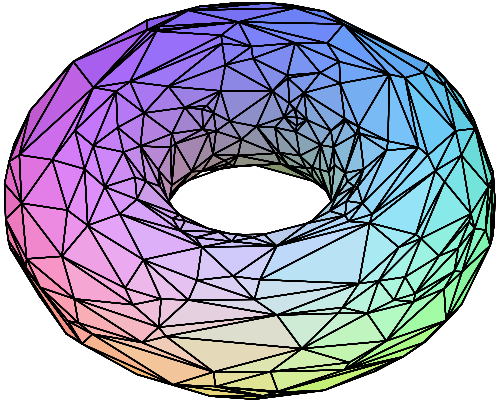

Figure 1. Random lattice with toroidal topology.

We then applied the reweighting method to compute, e.g., the specific heat, $C^{(i)}(\beta) = \beta^2 V \left( \langle e^2 \rangle - \langle e \rangle^2 \right)$, for each replica labeled by the superindex $(i)$, performed the replica average $C(\beta) = [C^{(i)}(\beta)] \equiv (1/20) \sum_i^{20} C^{(i)}(\beta)$, and finally determined the maximum, $C_{\max} = C(\beta_{C_{\max}})$. For the magnetic susceptibility, $\chi(\beta) = \beta V \left( [\langle m^2 \rangle - \langle m \rangle^2] \right)$, we followed exactly the same lines.

The proper replica average for the specific heat and susceptibility follows from the general rule that in the quenched case the free energy (and its derivatives) should be averaged [9]. For the (energetic) Binder parameter, usually defined for pure systems as $B(\beta) = 1 - \langle e^4 \rangle / 3 \langle e^2 \rangle^2$, the proper replica average is less clear to us. We have therefore studied three different definitions: $B_1(\beta) = 1 - [\langle e^4 \rangle / 3 \langle e^2 \rangle^2]$, $B_2(\beta) = 1 - [\langle e^4 \rangle] / 3 [\langle e^2 \rangle^2]$, and $B_3(\beta) = 1 - [\langle e^4 \rangle] / 3 [\langle e^2 \rangle]^2$. While in spin glass simulations [10] usually the analog of $B_3$ (with $e$ replaced by the overlap) is used, for a random bond Ising chain [11] a better scaling behaviour was observed for the analog of $B_1$ (with $e$ replaced by $m$).

## 3. RESULTS

Our estimates of the extrema of $C$, $\chi$, and $B_1$ for the various lattice sizes are collected in Table 1. The error bars are estimated by jack-knifing over the 20 replica. This takes into account both the statistical errors on each $C^{(i)}(\beta)$ and the fluctuations among the different replica. Already a first qualitative inspection of the data indicates that the first-order nature of the phase transition persists on quenched random lattices.

To make this statement more precise we performed a finite-size scaling (FSS) analysis. Assuming a first-order phase transition, we expect for large system sizes an asymptotic FSS behaviour of the form [12–14]

$$C_{\max} = a_C + b_C V + \ldots, \qquad (2)$$

$$\chi_{\max} = a_\chi + b_\chi V + \ldots, \qquad (3)$$

$$B_{i,\min} = a_{B_i} + b_{B_i}/V + \ldots, \qquad (4)$$

and

$$\beta_{C_{\max}} = \beta_t + c_C/V + \ldots, \qquad (5)$$

etc., where $\beta_t$ is the infinite volume transition point. The data for $C_{\max}$ and $\chi_{\max}$ shown in Fig. 2 is clearly consistent with this assumption. From least-square fits we obtained $a_C = 23.3(2.0)$, $b_C = 0.0659(30)$, with a goodness-of-fit parameter $Q = 0.16$ (corresponding to a chi-square per degree of freedom of 1.7), and $a_\chi = -0.70(43)$, $b_\chi = 0.0629(13)$, with $Q = 0.45$.

Also the data for the Binder parameter minima confirms the hypothesis of a first-order phase transition. Here the least-square fits gave $a_{B_1} = 0.6240(20)$, $b_{B_1} = -18.8(1.4)$, $Q = 0.17$, $a_{B_2} = 0.6236(22)$, $b_{B_2} = -18.5(1.4)$, $Q = 0.47$, and $a_{B_3} = 0.61125(68)$, $b_{B_3} = -16.45(71)$, $Q = 0.55$. Notice the much higher accuracy of $B_3$.

Our data for the pseudo-transition points and the corresponding fits through all data points are shown in Fig. 3. The resulting estimates for $\beta_t$ are $0.83360(14)$ from $C_{\max}$ ($Q = 0.51$), $0.83365(14)$ from $\chi_{\max}$ ($Q = 0.47$), and $0.83371(14)$ from $B_{1,\min}$ ($Q = 0.40$). On the scale of Fig. 3 the data points for $\beta_{B_2,\min}$ and $\beta_{B_3,\min}$ could hardly be disentangled from $\beta_{B_1,\min}$ and are therefore omitted. The results for $\beta_t$ are 0.83350(13) from $B_{2,\min}$ ($Q = 0.25$), and 0.83362(13) from $B_{3,\min}$ ($Q = 0.23$). By taking the average of these estimates we finally obtain

$$\beta_t = 0.83362 \pm 0.00013. \qquad (6)$$

Table 1
Extrema of the specific heat ($C_{\max}$), the susceptibility ($\chi_{\max}$), and the Binder parameter ($B_{1,\min}$), together with the corresponding pseudo-critical couplings.

| $V$ | $\beta_{C_{\max}}$ | $C_{\max}$ | $\beta_{\chi_{\max}}$ | $\chi_{\max}$ | $\beta_{B_{1,\min}}$ | $B_{1,\min}$ |
|---|---|---|---|---|---|---|
| 250  | 0.82500(44) | 33.15(45) | 0.82404(46) | 14.96(20)  | 0.81872(48) | 0.5662(11) |
| 500  | 0.82946(35) | 55.51(93) | 0.82907(34) | 31.09(56)  | 0.82655(34) | 0.5875(13) |
| 750  | 0.83087(23) | 76.1(2.0) | 0.83065(24) | 47.7(1.3)  | 0.82901(24) | 0.5960(18) |
| 1000 | 0.83112(31) | 90.4(2.6) | 0.83095(31) | 61.0(1.8)  | 0.82972(32) | 0.6044(17) |
| 2000 | 0.83232(22) | 144.8(9.0)| 0.83225(21) | 114.8(7.7) | 0.83164(21) | 0.6180(31) |
| 3000 | 0.83300(16) | 216(11)   | 0.83297(16) | 185.1(9.9) | 0.83257(16) | 0.6190(25) |

Notice that this value is very close to the exactly known transition point of the 8-state Potts model on a triangular lattice ($\beta_t^{\text{triang.}} = 0.85666\ldots$) [3].

Finally we show in Fig. 3 the "ratio-of-weights" definition of pseudo-transition points, $\beta_W$, which are expected to approach $\beta_t$ exponentially fast with increasing lattice size [15]. Basically the idea is to reweight the energy histograms to a point $\beta_W$ where the weights of the ordered and disordered phase are in a ratio $q : 1$. As in earlier studies for regular square lattices [14,15], we find also here that the $\beta_W$ are quite accurate estimates of $\beta_t$ already for very small system sizes.

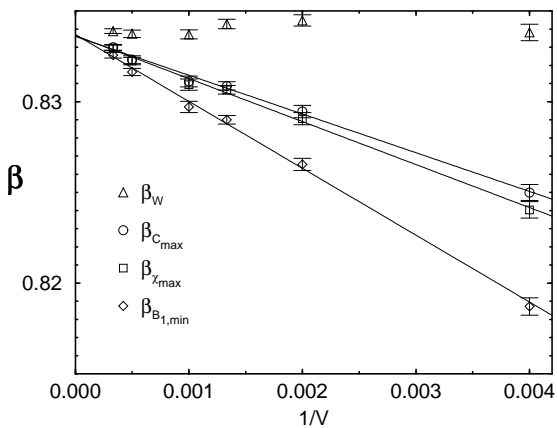

Figure 3. FSS of pseudo-transition points.

## 4. CONCLUSIONS

Summarizing, we have obtained clear numerical evidence for a first-order phase transition in the 8-state Potts model on quenched random lattices of Voronoi/Delaunay type. We can savely exclude a cross-over to a continuous transition as was observed for a certain type of quenched bond randomness on square lattices [4] and for the annealed disorder of dynamically triangulated surfaces [5].

This conclusion is based on the FSS behaviour of standard thermodynamic observables. We are currently extending the analysis to quantities that are directly related to the probability distri-butions of the energy or magnetization, such as the interface tension and the briefly mentioned "ratio-of-weights" definition of pseudo-transition points. Details of this study, which is based on a much larger set of 128 replica, will be presented elsewhere [16].

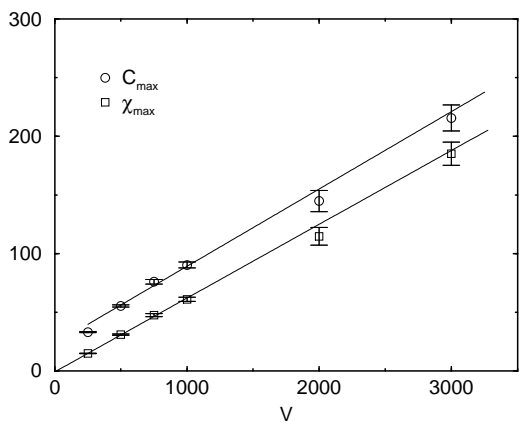

Figure 2. FSS of specific-heat and susceptibility maxima.

**ACKNOWLEGDEMENTS**

WJ would like to thank D.P. Landau for useful discussions and communicating the results of Ref. [4] prior to publication.

**REFERENCES**

1. A.B. Harris, J. Phys. C7 (1974) 1671; Y. Imry and S.-k. Ma, Phys. Rev. Lett. 35 (1975) 1388; A. Aharony, Phys. Rev. B18 (1978) 3318; G. Grinstein and S.-k. Ma, Phys. Rev. Lett. 49 (1982) 684; A.N. Berker, Phys. Rev. B29 (1984) 5293.
2. A.N. Berker, Phys. Rev. B29 (1984) 5293; K. Hui and A.N. Berker, Phys. Rev. Lett. 62 (1989) 2507; 63 (1989) 2433(E); A. Aizenman and J. Wehr, Phys. Rev. Lett. 62 (1989) 2503.
3. F.Y. Wu, Rev. Mod. Phys. 54 (1982) 235; 55 (1983) 315(E).
4. S. Chen, A.M. Ferrenberg, and D.P. Landau, Phys. Rev. Lett. 69 (1992) 1213 (1992); Phys. Rev. E52 (1995) 1377.
5. J. Ambjørn, G. Thorleifsson, and M. Wexler, Nucl. Phys. B439 (1995) 187.
6. W. Janke and R. Villanova, Mainz preprint KOMA-95-64 (June 1995), hep-lat/9507009.
7. C. Itzykson and J.-M. Drouffe, *Statistical Field Theory* (Cambridge University Press, Cambridge, 1989), Vol. 2; N.H. Christ, R. Friedberg, and T.D. Lee, Nucl. Phys. B202 (1982) 89; Nucl. Phys. B210 [FS6] (1982) 310, 337; R. Friedberg and H.-C. Ren, Nucl. Phys. B235 [FS11] (1984) 310.
8. W. Janke, M. Katoot, and R. Villanova, Phys. Lett. B315 (1993) 412; Phys. Rev. B49 (1994) 9644.
9. K. Binder and A.P. Young, Rev. Mod. Phys. 58 (1986) 801.
10. E. Marinari, G. Parisi, and F. Ritort, Roma preprint (1993), cond-mat/9310041.
11. A. Crisanti and H. Rieger, J. Stat. Phys. 77 (1994) 1087.
12. C. Borgs and R. Kotecky, J. Stat. Phys. 61 (1990) 79; Phys. Rev. Lett. 68 (1992) 1734; C. Borgs, R. Kotecky, and S. Miracle-Solé, J. Stat. Phys. 62 (1991) 529; J. Lee and J.M. Kosterlitz, Phys. Rev. B30 (1991) 3265; W. Janke, Phys. Rev. B47 (1993) 14757.
13. For a general overview see, e.g., K. Binder, Rep. Prog. Phys. 50 (1987) 783; or the articles in *Dynamics of First Order Phase Transitions*, eds. H.J. Herrmann, W. Janke, and F. Karsch (World Scientific, Singapore, 1992).
14. W. Janke, in *Computer Simulations in Condensed Matter Physics VII*, eds. D.P. Landau, K.K. Mon, and H.B. Schüttler (Springer Verlag, Heidelberg, Berlin, 1994), p. 29; and references therein.
15. C. Borgs and W. Janke, Phys. Rev. Lett. 68 (1992) 1738.
16. W. Janke and R. Villanova, in preparation.